\documentclass[final,3p,times,twocolumn]{style}

\usepackage{graphicx}
\usepackage{amsmath}
\usepackage{stmaryrd}
\usepackage[english]{babel}
\usepackage{lineno}
\usepackage{url}
\usepackage{xcolor}
\usepackage{soul}

\setcitestyle{semicolon,aysep={,}}

% ...    

\begin{document}

\begin{frontmatter}




\title{Complexity in patterns of racial segregation}


\author[add1]{Tomasz F. Stepinski}
\author[add1,add2]{Anna Dmowska}

\address[add1]{Space Informatics Lab, University of Cincinnati,  Cincinnati, USA, OH 45221-0131, USA}
\address[add2]{Institute of Geoecology and Geoinformation, Adam Mickiewicz University, Poznan, Poland}
\cortext[cor2]{Corresponding author. {\it Email address:} stepintz{@}uc.edu}

\begin{abstract}
%
Cities are complex systems, their complexity manifests itself through fractality of their spatial structures and by power law distributions (scaling) of multiple urban attributes. Here we report on the previously unreported manifestation of urban complexity -- scaling in patterns of residential racial segregation. A high-resolution racial grid of a city is segmented into racial enclaves which are patches of stationary racial composition. Empirical PDFs of patch areas and population counts in 41 US cities were analyzed to reveal that these variables have distributions which are either power laws or approximate power laws. Power law holds for a pool of all patches, for patches from individual cities, and patches restricted to specific racial types. The average value of the exponent is 1.64/1.68 for area/population in 1990 and 1.70/1.74 in 2010. The values of exponents for type-specific patches vary, but variations had decreased from 1990 to 2010. We have also performed a multifractal analysis of patterns formed by racial patches and found that these patterns are monofractal with average values of fractal dimensions in the 0.94-1.81 range depending on racial types and the year of analysis. Power law distribution of racial patch sizes and a fractal character of racial patterns present observable and quantifiable constraints on models of racial segregation. We argue that growth by preferential attachment is a plausible mechanism leading to observed patterns of segregation.
%
\end{abstract}

\begin{keyword}

Urban complexity \sep power law scaling \sep multifractal analysis \sep racial segregation \sep preferential attachment


\end{keyword}

\end{frontmatter}


\section{Introduction}
Cities are complex systems and their structures emerge from behaviors of their many elementary constituents (social, economic, etc.) interacting with each other on a variety of spatial and temporal scales. The notion of a city as a complex system is supported by empirical evidence. The evidence can be divided into spatial and aspatial. 

Spatial evidence stems from the fact that cities' spatial structures are best described in terms of fractal geometry \citep{Batty1994,DeKeersmaecker2003,Chen2013}.
Urban morphologies display statistical self-similarity \citep{Batty1989,SAMBROOK2001} which indicates the presence of a hidden process that operates at different scales. Various types of data were used as observations of urban spatial structures. The most commonly used is the data of built-up areas which originates from remotely-sensed images \citep{Murcio2011,Chen2013,Song2019}. Maps of population density are also used \citep{ChenFeng2017}, as well as remotely-sensed images of night-lights \citep{Ozik2005} and street intersection point patterns \citep{Murcio2015}. Analysis of various datasets revealed that urban structure has a multifractal rather than a monofractal nature \citep{Appleby1996,Benguigui2004,Cavaihes2010,Ariza-Villaverde2013,Murcio2015} with the degree of multifractality depending on the modality of the data \citep{Kia2020}.
Additional spatial evidence comes from the time-evolution of cities that shows spreading, leapfrogging \citep{Benguigui2004}, and hotspots behavior which suggest a complex, non-linear mechanism.

\begin{figure*}[t]
	\includegraphics[width=15cm]{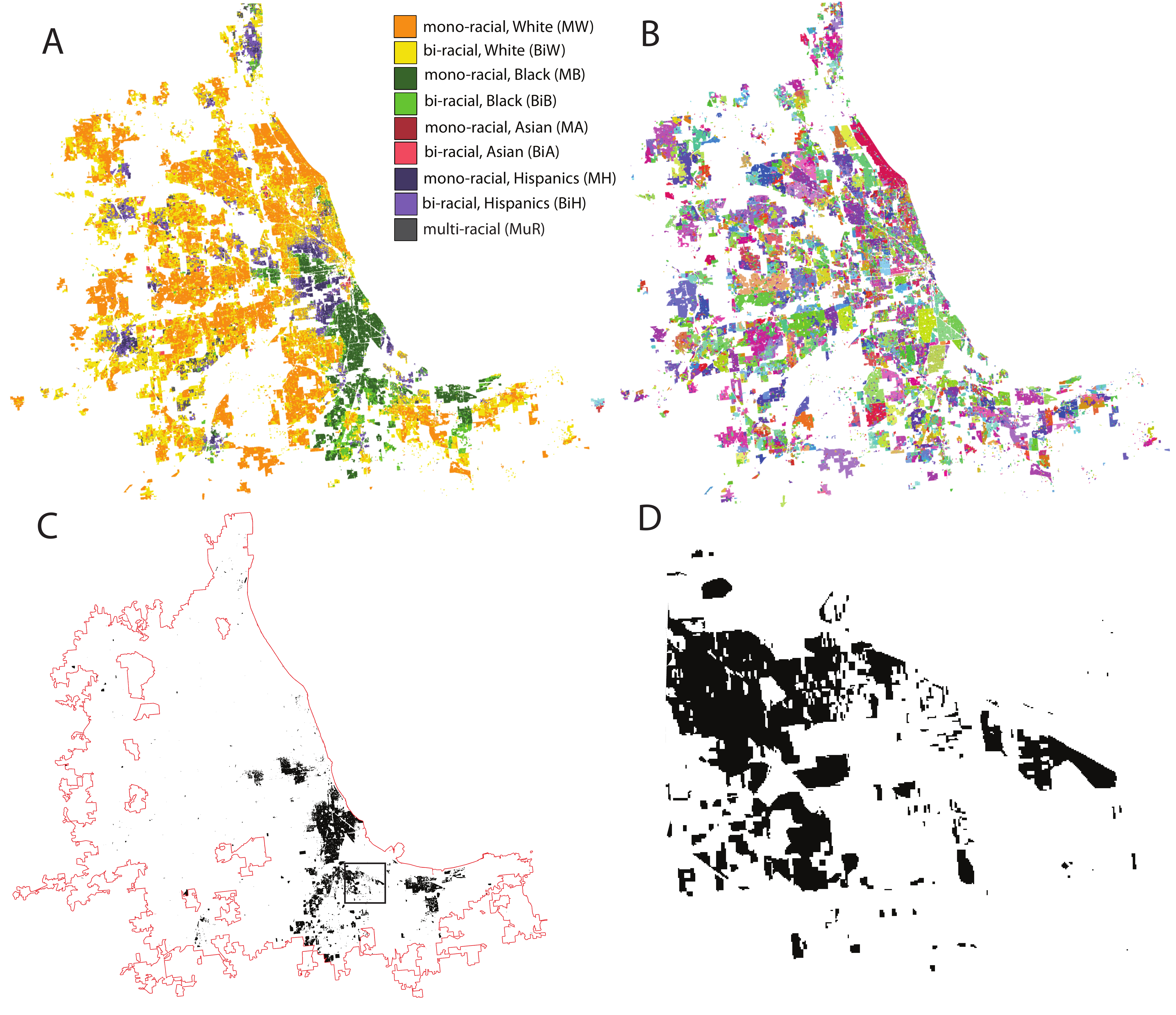}
\caption{Racial urban pattern in Chicago, IL urban area in 2010. (A) Map of racial patches color-coded for their types. (B) Map of over 14,000 racial patches randomly-colored. (C) Map of mono-racial Black patches. (D) Enlargement of the area indicated by a rectangle in panel (C). White color indicates uninhabited areas. 
}
\label{fig1}
\end{figure*}

Aspatial evidence comes from the observation that many urban quantities obey a power law distribution. Aspatial evidence may be divided into inter-urban and intra-urban. The inter-urban analysis reveals that many quantities characterizing cities as a whole scale with the city size \citep{Bettencourt2007,Batty2008,Bettencourt2013}. Perhaps the best know inter-urban scaling property is captured by Zipf's law for city size distribution (for a survey of literature on Zipf's law see \citep{Arshad2018,Cottineau2017}. Empirical data shows that the probability distribution of urban populations follows the power law with exponent $\alpha \approx 2$. The value of $\alpha$ stays close to the value of 2 regardless of whether cities are collected from a single country, a continent, or the entire world \citep{Fang2017,Oliviera2018}. In \citep{Pumain2006,Gomez2017} it was demonstrated that per capita rates of many different urban quantities (divided into types like employment, innovation, crime, educational attainment, and infectious diseases) also scale with the population size.

Urban analysis on the intra-urban scale reveals the complexity of the spatial organization of an individual city. In \citep{Lammar2006} and \citep{Strano2012} distribution of urban blocks were studied. Blocks were defined as segments of the planar urban structure surrounded by street segments (not to be mistaken with US census blocks which are statistical agglomerative areas). They found that block area distribution is a power law with $\alpha \approx 2$. In \citep{Fialkowski2008, Bitner2009} distributions of cadastral parcels in large North American and Australian cities were studied. Parcels were divided into inhabited and uninhabited. It was demonstrated that inhabited parcel size distribution is a power law with $\alpha \approx 2$, while uninhabited parcel size distribution is a power law with $\alpha \approx 1$. \citet{Riascos2017} analyzed the size distribution of lots occupied by buildings and uninhabited lots using data from the OpenStreetMaps (OSM). Building lots are substructures of blocks studied in \citep{Lammar2006, Strano2012}. Using data for Berlin, Hamburg, London, and New York, he has found that uninhabited lots size distribution is a power law with $\alpha \approx 1$, but building lots size distribution is a power law with $\alpha \approx 3$. \citet{Carvalho2008} shown that distributions of building heights, areas, and volumes in Greater London are power law.

Urban populations in the United States are multiracial, but also, to a large degree, segregated \citep{Logan2004, Parisi2011, Lee2014}. Increasingly, major European cities are also becoming multiracial and segregated \citep{Andersson2018,Sturgis2014}. Racial segregation is an important social component of city organization. Is the scaling properties found in previously studied urban quantities extent to racial segregation? 
To address this question we segmented a high-resolution SocScape racial grid \citep{Dmowska2017} (a categorical-valued matrix-form dataset) covering a given city into racial patches. A racial patch is a contiguous set of same-type grid cells and is characterized by a stationary racial composition. Patches are mono-racial, bi-racial, and some of them can be multi-racial. Patches offer the best partitioning of races given that different racial subpopulations have overlapping distributions.

Fig.~1 illustrates racial patches as an intro-urban categorical quantity using the urban area of Chicago, IL as an example. In panel A, patches are colored by nine different types described in detail in the next section; boundaries between different patches are not clearly visible due to the scale of the map. In panel B, each of $\sim$ 14,000 patches is assigned a random color. In panel C, only a single type of patches (MB) is mapped. Panel D shows an enlargement of the areas indicated by a rectangle in panel C; at this magnification orders of magnitude differences in patch sizes are clearly visible suggesting a possibility of a heavy-tailed, possibly a power law-like, distribution of patch sizes. In addition, patterns shown in Fig.~1C, and D appear to be fractal. To quantify these visual insights we calculate patch size distributions for all patches as well as for type-based subsamples of patches using 1990 and 2010 gridded data for the set of 41 major U.S. cities. We define the ``patch size" as either patch's area or its population count. We also perform multifractal analysis of patterns formed by the patches. 

\section{Data and Method}
It is important to note that it is not possible to segment a city into racially homogeneous patches. Instead, we segment cities into patches of homogeneous racial character (stationary racial composition). For such segmentation, we use a high-resolution racial grid for the conterminous U.S. \citep{Dmowska2017} accessible through the SocScape web application (http://sil.uc.edu/webapps/socscape usa/). This product is a gridded categorical map with each inhabited, $30{\rm m}\times 30{\rm m}$, cell labeled by one of 13 types on the basis of local racial diversity and the dominant race. Note that streets have been filtered out from diversity grids, but larger roads are present and appear as uninhabited areas.

The cell's type is based on racial diversity and a dominant race of its population.  The degree of diversity is quantified by the entropy $E$ of local racial composition and the percentage of cell's dominant subpopulation \citep{Dmowska2014}. Based on these input cells are classified into low diversity (mono-racial), medium diversity (bi-racial), and high diversity (multiracial). Mono-racial and bi-racial cells are further classified on the basis of the dominant race, White, Black, Hispanic, Asian, American Indians, and Others resulting in 13 possible types. Fig.~1A shows a 2010 map of these cell types segmented into patches for Chicago urban area, we use abbreviation labels of types listed in the legend of this figure throughout this paper.  

We use connected components labeling algorithm \citep{Rosenfeld1966} as implemented in \citep{Netzel2013} to segment a racial grid. We use a 4-connectivity (rook contiguity) variant of an algorithm. An output of an algorithm is a set of patches (spatially contiguous regions of same-type cells). Fig.~1B shows a result of segmenting a racial map of Chicago. Each patch has its unique identification (ID) number which in Fig.~1B is indicated by random color. In addition to its ID number, a patch is also characterized by its type (one of 13), area, and population count. For our analysis, we select patches fulfilling all of the following conditions: area $\ge$ 10 cells, population $\ge$ 10 people, population density $\ge$ 0.7 people/cell. This eliminates uninhabited patches, as well as patches which are too scarcely populated to be considered urban and patches which are too small to consider as viable urban units.

\subsection{Fitting and testing the power law distribution}
We first assume that patch size distribution $N(x)$ is a power law and estimate its exponent. The patch size, $x$, is either its area or its population count, and $N(x)$ denotes the number of patches of size $x$. The analysis follows a procedure described in \citep{Newman2005,Clauset2009}. The power law distribution is described by a probability density $p(x)$ such that
\begin{equation}
p(x) = \frac{N(x)}{N_0} = C x^{-\alpha} = \frac{\alpha -1}{x_{min}} \left(\frac{x}{x_{min}}\right)^{-\alpha}
\end{equation}
\noindent where $N_0$ is the total number of patches and $C$ is a normalization constant present to assure that $\int p(x) dx$ over all possible values of $x$ equals 1. For this to be possible in power laws with positive values of exponent $\alpha$, there must be a lower bound, $x_{min}$ to the power law behavior. Thus, fitting the power law model to the data is tantamount to finding best-fit values of $\alpha$ and  $x_{min}$. However, because we are interested in finding the most general rule for the size distribution of racial patches in all U.S. cities, we keep the value of $x_{min}$ constant and equal to 30 cells for area distribution and 100 people for population distribution. These values reflect our opinion on what minimum area/population size constitutes a patch. With the value of the lower bound  $x_{min}$ fixed, the value of $\alpha$ is estimated using the formula derived from the maximum likelihood estimator (MLE) \citep{Newman2005,Clauset2009}.

\begin{equation}
\alpha = 1 + n \left( \sum_{i=1}^{n} \ln \frac{x_i}{x_{min}}\right)^{-1}
\end{equation}

\noindent where $n$ is the number of patches and $x_i$ is the size of the $i$-th patch.

\begin{figure*}[t]
	\includegraphics[width=16cm]{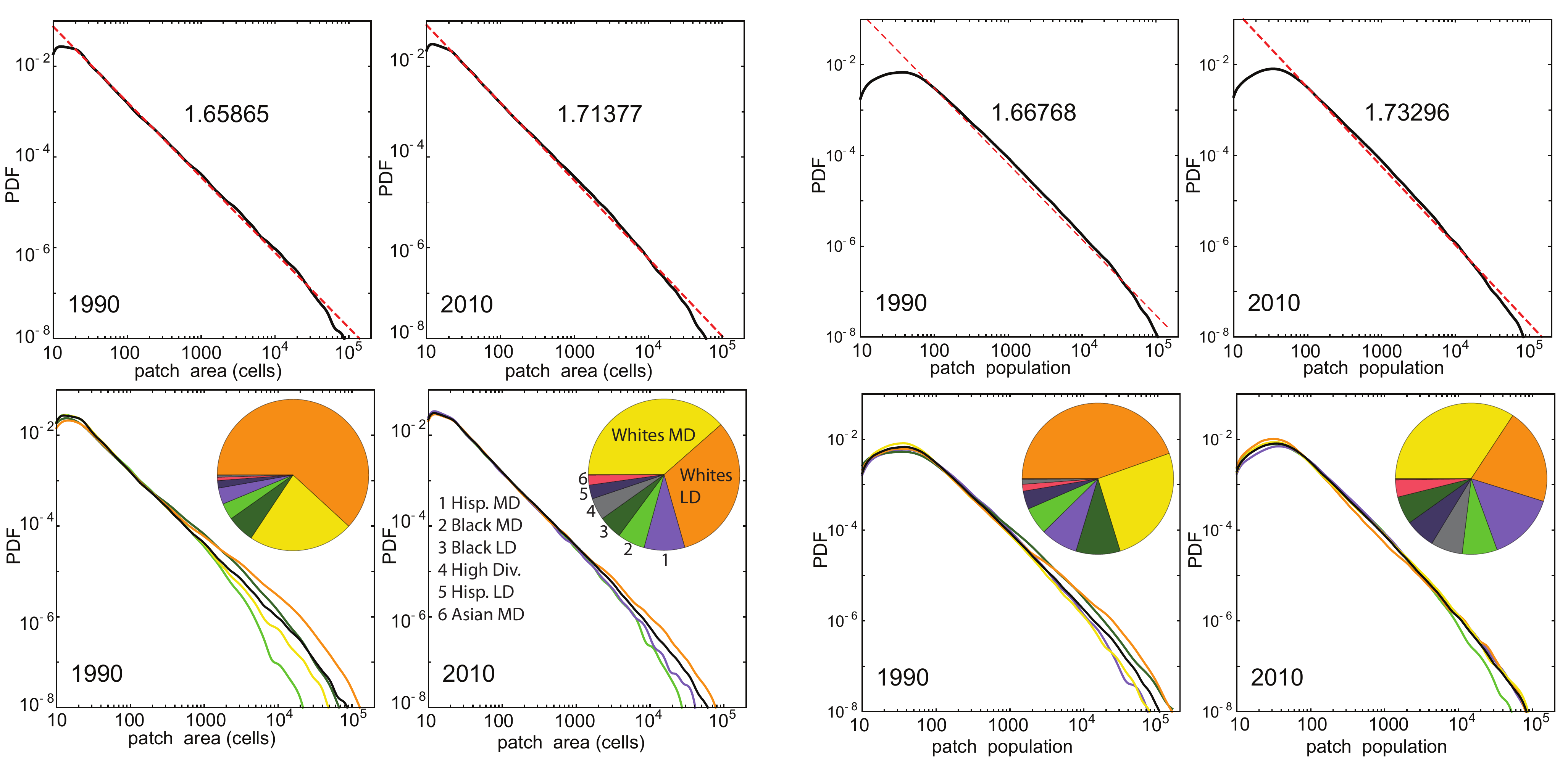}
\caption{Kernel density estimations of
 PDFs for the pool of patches from all 41 cities in 1990 and 2010. The four panels on the left side pertain to an area distribution and the four panels on the right side pertain to population distribution. In each group, upper panels show an empirical PDF (solid black curve) and the fitted power law (dashed red curve); the number above the curve is the fitted exponent. Lower panels show a comparison between a PDF constructed from all patches with PDFs constructed from single-type patches corresponding to the top four patches types. Pie charts illustrate the relative composition of patches types (by area for left panels and by population for right panels). Abbreviations: Hisp. -- Hispanics, Div. -- diversity, LD -- low diversity, MD -- medium diversity.}
\label{fig2}
\end{figure*}

In \citep{Broido2019} it was pointed out that most reports of the power law behavior in various systems (including those in urban structures) are questionable because the power law form of the variable distribution was not statistically tested. Thus, we test the power law hypothesis using a bootstrapping method \citep{Clauset2009}. The divergence between the empirical complementary CDF (cCDF) and our model cCDF is measured by the Kolmogorov-Smirnov (KS) distance $D^{\star}=\max_{x \geq x_{min}} |F(x)-P(x)|$, where $F(x)$ is the empirical cCDF and $P(x)$ is the model (fitted power law) cCDF. For comparison, we calculate a KS distance $D$ between the empirical cCDF constructed from values randomly drawn from our model and the cCDF of a model fitted to these values. We generated 1000 random sets of data resulting in 1000 values of $D$. The value $R =D^{\star}/\langle D \rangle$ is a ratio of the goodness-of-fit of our data to its assumed power law model to the goodness-of-fit
of data generated from our power law model to its own power law model. If $R \sim 1$ the power law hypothesis is justified, but if $R>>1$ the power law hypothesis is not justified. We like to use $R$ because it quantifies how far our model is from being statistically considered a truth hypothesis. The actual test uses the $p$-value which is the number of times $D>D^{\star}$ divided by the total number of random sets (1000); it provides only true/untrue answer to the hypothesis. We also indicate the result of the test based on the $p$-value.

\subsection{Fractal analysis}
A spatial structure of the entire city is a pattern formed by all patches (for example, a pattern shown in Fig.~1B). A spatial structure of a part of the city inhabited by a given racial type is a pattern formed by patches of this type (for example, a pattern shown in Fig.~1C). 

We perform a multifractal analysis of these patterns using the box-counting method \citep{Cheng1995}. An urban area is divided into a series of grids with different box sizes. For a grid with given box size, ($\epsilon$), a ``mass" (a number of raster cells in a box) is calculated and stored in the originating box thus transforming a grid to a numerical array. Dividing the mass of each box by the total mass of all urban elements in a region yields a probability distribution $\{p_i\}=\{p_1,\dots p_n\}$, where $n$ is the number of nonempty boxes. We calculate Renyi's generalized dimensions $D_q$ to reveal a multifractal description of a pattern using a method described in \citep{Murcio2015,Kia2020}. A value of $D_q$ is defined as
\begin{equation}\label{Dqfinal}
D_q =
\begin{cases}
{\displaystyle\frac{1}{q-1} \lim_{\epsilon\to 0}} \frac{\log M_q(\epsilon)}{\log\epsilon}  & \text{if}\ \  q \neq 1 \\
\\
-{\displaystyle \lim_{\epsilon\to 0}} \frac{-\sum_{i=1}^n p_i(\epsilon) \log p_i(\epsilon) }{\log\epsilon}  & \text{if}\ \  q=1 
\end{cases}
\end{equation}
\noindent where $M_q (\epsilon) = \sum p_i(\epsilon)^q$ is a statistical moment of order $q$; $\infty \le q \le \infty$.

If all moments scale at the same rate the pattern is monofractal, if they
scale with different rates the pattern is multifractal, and if they don't
scale the pattern is nonfractal.

\section{Results}
We applied the methodology described above to the segments' area and population distributions in 41 major U.S. cities listed in Table 1. US Census 2010 Urban Areas boundaries are used to delineate cities' spatial extents. For each city, we calculated the best-fit power law exponent and its error (see Table 1), tested the power law hypothesis (see Table 2), and constructed a kernel density estimation of empirical PDF for visual comparison with the PDF of the fitted power law model. 

Fig.~2 shows the results for a pool of patches from all 41 cities; there are 91,768 patches in such a pool in 1990 and 183,229 patches in 2010. The large difference in the number of patches between 1990 and 2010 is due to the U.S. Census Bureau dividing 1990 blocks into smaller 2010 blocks in order to separate inhibited and uninhabited parts of blocks. This figure has eight panels organized into two groups. The left group of four panels pertains to an area distribution and the right group of four panels pertains to population distribution. In each group, the two upper panels show empirical PDFs constructed from all patches in 1990 and 2010 together with their power law fits. Two lower panels show 1990 and 2010 comparisons of empirical PDFs constructed from all patches (black curves) with PDFs constructed from single-type subsets of patches (colored curves).

Table 1 summarizes the values of exponents fitted to distributions of patch area ($\alpha_a$) and patch population ($\alpha_p$) in each of the 41 cities in our sample. We also list estimated errors of exponents' values. The bottom of the table shows average values of $\alpha_a$ and $\alpha_p$ together with the values of their standard deviations. The average values of exponents are very similar to the values of exponents calculated from the pool of patches from all cities.

Values of exponents $\alpha_a$ have increased during the 1990-2010 period for 33 cities. For one city (Portland, OR) the value of exponent has decreased, and for seven cities the values did not change within the error bars. Values of exponents $\alpha_p$ have increased during the 1990-2010 period for 33 cities. For two cities (Portland, OR and Phoenix, AZ) values of exponents have decreased, and for six cities they did not change within the error bars. In 1990 values of $\alpha_{p}-\alpha_{a}$ were positive for 23 cities, negative for 2 cities (Los Angeles, CA and New York, NY), and zero within error bars for 16 cities. In 2010 values of $\alpha_p - \alpha_a$ were positive for 28 cities, negative for three cities (Los Angeles, CA, New York, NY, and San Jose, CA), and zero within error bars for 10 cities. 

Table 2 summarizes the results of testing the power law assumption. Values of $R_a$ pertain to testing power law hypothesis for distributions of patch sizes and values of $R_p$ pertain to testing power law hypothesis for distributions of patch population. Distributions with $p \le 0.1$ (conservative test) are indicated by superscript ${\ast \ast}$ and distributions with $p \le 0.05$ (lenient test) are indicated by superscript ${\ast}$.

In 1990, the power law distribution of patch areas cannot be ruled out for 22 cities (19 at the $p\le 0.1$ level) and the power law distribution of patch populations cannot be ruled out for 17 cities (14 at the $p\le 0.1$ level). In 2010, the power law distribution of patch areas cannot be ruled out for 26 cities (18 at the $p\le 0.1$ level) and the power law distribution of patch populations cannot be ruled out for 12 cities (10 at the $p\le 0.1$ level). Thus, in general, distributions of patch areas appear to be closer to the real power law than distributions of patch populations.

\begin{table*}[]
\caption{\textbf{{Exponents of fitted power laws}}}
{\footnotesize \vspace*{3mm} \hspace*{2mm}
\begin{tabular}{p{1.4cm}|p{0.5cm} p{0.55cm} p{0.55cm} p{0.5cm} p{0.55cm} p{0.8cm} || p{1.4cm}| p{0.6cm} p{0.55cm} p{0.55cm} p{0.6cm} p{0.55cm} p{0.7cm}}
\hline
 & \multicolumn{3}{|c}{1990} &  \multicolumn{3}{c||}{2010} &  & \multicolumn{3}{c}{1990} &  \multicolumn{3}{c}{2010}  \\
city name & patches & $\alpha_{a}$ & $\alpha_{p}$ & patches & $\alpha_{a}$ & $\alpha_{p}$ & city name &patches & $\alpha_{a}$ & $\alpha_{p}$ & patches & $\alpha_{a}$ & $\alpha_{p}$ \\
\hline
Atlanta & 1646 & 1.56(1) & 1.61(2) & 5070 & 1.65(1) & 1.68(1) & New York  & 11821 & 1.71(1) & 1.61(1) & 19129 & 1.74(1) & 1.68(0) \\
Austin & 903 & 1.74(2) & 1.74(2) & 2053 & 1.72(2) & 1.72(2) & Orlando & 1097 & 1.62(2) & 1.69(2) & 3074 & 1.75(1) & 1.79(1) \\
Baltimore & 1225 & 1.53(2) & 1.56(2) & 2691 & 1.61(1) & 1.65(1) & Philadelphia & 3287 & 1.60(1) & 1.63(1) & 7210 & 1.68(1) & 1.71(1)\\
Boston & 2072 & 1.57(1) & 1.58(1) & 4851 & 1.70(1) & 1.71(1) & Phoenix & 2132 & 1.78(2) & 1.85(2) & 5498 & 1.79(1) & 1.79(1)\\
Charlotte &558 &	1.66(3) &1.75(3) &1802 & 1.63(1) &1.74(2) & Pittsburgh &1026 &1.55(2) &	1.64(2) &1887 &1.6692) &1.73(2) \\
Chicago &	5819 &	1.66(1) &	1.68(1) &	14668 &	1.75(1) &	1.78(1) & Portland &	1208 &	1.80(2) &	1.82(2) &	3002 &	1.72(1) &	1.76(1) \\
Cincinnati &	664 &	1.46(2) &	1.54(2) &	1477 &	1.54(1) &	1.63(2) & Providence &	707 &	1.74(3) &	1.76(3) &	1562 &	1.80(2)	& 1.82(2)\\
Cleveland &	734 &	1.50(2) &	1.54(2) &	1524 &	1.61(2) &	1.66(2) & Riverside &	1472 &	1.69(2) &	1.77(2) &	2404 &	1.71(1) &	1.76(2) \\
Columbus &	625 &	1.57(2) &	1.64(3) &	1972 &	1.63(1) &	1.71(2) & Sacramento &	1313 &	1.77(2) &	1.80(2) &	2979 &	1.79(1) &	1.85(2)\\
Dallas &	4278 &	1.69(1) &	1.69(1) &	9722 &	1.74(1) &	1.76(1) & St. Louis &	1255 &	1.61(2) &	1.66(2) &	2834 &	1.66(1) &	1.75(1)\\
Denver &	2227 &	1.72(2) &	1.75(2) &	4262 &	1.76(1) &	1.79(1) & Salt Lake C. &	462 &	1.56(3) &	1.56(3) &	1398 &	1.71(2) &	1.74(2) \\
Detroit &	1851 &	1.64(1) &	1.67(2) &	4575 &	1.72(1) &	1.76(1) & San Antonio &	1588 &	1.72(2) &	1.75(2) &	2247 &	1.72(2) &	1.73(2) \\
Houston &	4097 &	1.71(1) &	1.73(1) &	8007 &	1.73(1) &	1.74(1) & San Diego &	2003 &	1.63(1) &	1.66(1) &	3555 &	1.70(1) &	1.74(1)\\
Indianapolis &	627 &	1.56(2) &	1.62(2) &	2144 &	1.66(1) &	1.73(2) & S. Francisco &	3318 &	1.74(1) &	1.74(1) &	4638 &	1.79(1) &	1.78(1) \\
Jacksonville &	889 &	1.60(2) &	1.66(2) &	1814 &	1.67(2) &	1.72(2) & San Jose &	1501 &	1.69(2) &	1.69(2) &	2134 &	1.80(2) &	1.76(2)\\
Kansas City &	1152 &	1.63(2) &	1.69(2) &	2813 &	1.69(1) &	1.76(1) & Seattle &	2130 &	1.65(1) &	1.71(2) &	4924 &	1.72(1) &	1.78(1)\\
Las Vegas &	648 &	1.70(3) &	1.71(3) &	2544 &	1.74(1) &	1.73(1) & Tampa &	2340 &	1.69(1) &	1.73(2) &	4980 &	1.73(1) &	1.78(1) \\
Los Angeles &	10308 &	1.72(1) &	1.70(1) &	12640 &	1.78(1) &	1.73(1) & Virginia B. &	1488 &	1.56(1) &	1.62(2) &	2691 &	1.64(1) &	1.69(1)\\
Memphis &	684 &	1.59(2) &	1.64(2) &	1448 &	1.66(2) &	1.71(2) & Washington &	3068 &	1.56(1) &	1.58(1) &	6033 &	1.65(1) &	1.67(1)\\
Miami &	5483 &	1.66(1) &	1.68(1) &	8701 &	1.70(1) &	1.71(1) & Milwaukee &	742 &	1.58(2) &	1.66(2) &	1703 &	1.71(2) &	1.75(2)\\
Minneapolis &	1320 &	1.65(2) &	1.71(2) &	4569 &	1.72(1) &	1.79(1) &  & &	 &	 &	 &	 &	\\
\hline
average &	 &	1.64 &	1.68 & &	1.70 &	1.74 & all cities & 91768&	1.66(0) &	1.67(0) &183229	 &	1.71(0) & 1.73(0)	\\
 &	 &	$\pm$0.08 &	$\pm$0.07 & &	$\pm$0.06 &	$\pm$0.05 &  & &	 &	 &	 &	 & 	\\
\hline
\multicolumn{14}{l}{%
  \begin{minipage}{15.5cm}%
{\vspace{1mm} \footnotesize {Values of exponents are shown with uncertainties calculated using the formula 3.2 in \citet{Clauset2009}. For example, for Atlanta, the value of $\alpha_{a}$ in 1990 is 1.56454 and the uncertainty is 0.0139148, resulting in the Table 1 entry 1.56(1). The value of exponent is rounded to 3 significant digits and the value of error is rounded to one significant digit. Abbreviations: patches -- number of area patches (number of population patches is slightly different), Salt Lake C. -- Salt lake City, S. Francisco -- San Francisco, Virginia B. -- Virginia Beach.  } }
  \end{minipage}%
}
\end{tabular}}
\label{exponents}
%
\end{table*}

\begin{table*}[]
\caption{\textbf{{Testing the power-law assumption}}}
{\footnotesize \vspace*{3mm} \hspace*{2mm}
\begin{tabular}{p{1.4cm}|p{0.5cm} p{0.55cm} p{0.55cm} p{0.5cm} p{0.55cm} p{0.8cm} || p{1.4cm}| p{0.6cm} p{0.55cm} p{0.55cm} p{0.6cm} p{0.55cm} p{0.7cm}}
\hline
 & \multicolumn{3}{|c}{1990} &  \multicolumn{3}{c||}{2010} &  & \multicolumn{3}{c}{1990} &  \multicolumn{3}{c}{2010}  \\
city name & patches & $R_{a}$ & $R_{p}$ & patches & $R_{a}$ & $R_{p}$ & city name &patches & $R_{a}$ & $R_{p}$ & patches & $R_{a}$ & $R_{p}$ \\
\hline
Atlanta & 1646 & 2.74 & 3.50 & 5070 & 2.30 & 4.48 & New York  & 11821 & 1.32$^{\ast}$ & 6.94 & 19129 & 1.44 & 6.30 \\
Austin & 903 & 1.35$^{\ast \ast}$ & 1.36$^{\ast \ast}$ & 2053 & 1.34$^{\ast \ast}$ & 2.07 & Orlando & 1097 & 1.01$^{\ast \ast}$ & 1.78 & 3074 & 1.36$^{\ast}$ & 1.41$^{\ast}$ \\
Baltimore & 1225 & 1.96 & 3.16 & 2691 & 2.53 & 3.70 & Philadelphia & 3287 & 1.77 & 3.23 & 7210 & 2.48 & 4.61\\
Boston & 2072 & 1.58 & 2.92 & 4851 & 1.55 & 4.15 & Phoenix & 2132 & 1.78 & 1.24$^{\ast \ast}$ & 5498 & 0.88$^{\ast \ast}$ & 1.98\\
Charlotte &558 &	1.18$^{\ast \ast}$ &1.59 &1802 & 2.01 &2.36 & Pittsburgh &1026 &1.08$^{\ast \ast}$ &	1.35$^{\ast \ast}$ &1887 & 1.54 & 2.04 \\
Chicago &	5819 &	1.18$^{\ast \ast}$ &	3.06 &	14668 &	1.23$^{\ast}$ & 3.61 & Portland &	1208 &	2.43 &	1.24$^{\ast \ast}$ &	3002 &	0.99$^{\ast \ast}$ &	1.89 \\
Cincinnati &	664 &	2.08 & 2.13 &	1477 & 2.52 &	3.18 & Providence &	707 &	0.79$^{\ast \ast}$ &	1.16$^{\ast \ast}$ &	1562 &	1.06$^{\ast \ast}$	& 0.71$^{\ast \ast}$\\
Cleveland &	734 &	1.59 & 2.11 &	1524 &	1.20$^{\ast \ast}$ &	1.60 & Riverside &	1472 &	2.07 &	1.28$^{\ast \ast}$ &	2404 &	0.93$^{\ast \ast}$ &	1.29$^{\ast \ast}$ \\
Columbus &	625 &	1.59 &	1.59 &	1972 &	1.42$^{\ast}$ &	2.05 & Sacramento &	1313 &	0.84$^{\ast \ast}$ &	1.35$^{\ast \ast}$ &	2979 &	0.60$^{\ast \ast}$ &	0.60$^{\ast \ast}$\\
Dallas &	4278 &	1.55 &	3.34 &	9722 &	1.84 &	4.26 & St. Louis &	1255 &	1.15$^{\ast \ast}$ &	1.10$^{\ast \ast}$ &	2834 &	1.12$^{\ast \ast}$ &	1.22$^{\ast \ast}$\\
Denver &	2227 &	1.27$^{\ast \ast}$ &	0.80$^{\ast \ast}$ &	4262 &	0.77$^{\ast \ast}$ &	1.68 & Salt Lake C. &	462 &	1.91 &	1.39$^{\ast}$ &	1398 &	1.39$^{\ast}$ &	0.74$^{\ast \ast}$ \\
Detroit &	1851 &	1.39$^{\ast}$ &	1.54 &	4575 &	0.91$^{\ast \ast}$ &	1.14$^{\ast \ast}$ & San Antonio &	1588 &	0.71$^{\ast \ast}$ &	1.26$^{\ast \ast}$ &	2247 &	1.65 &	0.94$^{\ast \ast}$ \\
Houston &	4097 &	1.81 &	2.46 &	8007 &	1.29$^{\ast}$ &	4.60 & San Diego &	2003 &	1.30$^{\ast \ast}$ &	1.79 &	3555 &	0.83$^{\ast \ast}$ &	1.10$^{\ast \ast}$\\
Indianapolis &	627 &	1.25$^{\ast \ast}$ &	2.07 &	2144 &	1.60 &	2.80 & S. Francisco &	3318 &	0.87$^{\ast \ast}$ &	2.73 &	4638 &	1.34$^{\ast}$ &	2.25 \\
Jacksonville &	889 &	1.30$^{\ast \ast}$ &	1.89 &	1814 &	1.42$^{\ast}$ &	1.72 & San Jose &	1501 &	1.25$^{\ast \ast}$ &	1.96 &	2134 &	0.77$^{\ast \ast}$ &	1.71\\
Kansas City &	1152 &	1.18$^{\ast \ast}$ &	1.39$^{\ast}$ &	2813 &	1.20$^{\ast \ast}$ &	2.54 & Seattle &	2130 &	1.46$^{\ast}$ &	0.98$^{\ast \ast}$ &	4924 &	0.77$^{\ast \ast}$ &	2.07\\
Las Vegas &	648 &	0.84$^{\ast \ast}$ &	1.40$^{\ast}$ &	2544 & 0.90$^{\ast \ast}$ &	1.42$^{\ast}$ & Tampa &	2340 &	1.09$^{\ast \ast}$ &	1.28$^{\ast \ast}$ &	4980 &	1.40$^{\ast}$ & 2.37 \\
Los Angeles &	10308 &	1.63 &	4.12 &	12640 &	0.80$^{\ast \ast}$ &	3.79 & Virginia B. &	1488 &	3.24 & 4.48 &	2691 &	2.83 &	1.64\\
Memphis &	684 &	1.54 &	1.98 &	1448 &	1.08$^{\ast \ast}$ &	1.18$^{\ast \ast}$ & Washington &	3068 &	3.63 &	5.83 &	6033 &	2.96 &	4.47\\
Miami &	5483 &	1.77 &	2.48 &	8701 &	1.57 &	1.70 & Milwaukee &	742 &	1.12$^{\ast \ast}$ &	1.11$^{\ast \ast}$ &	1703 &	0.95$^{\ast \ast}$ &	1.20$^{\ast \ast}$\\
Minneapolis &	1320 &	1.54 &	1.20$^{\ast \ast}$ &	4569 &	1.61 &	2.89 & all cities  &  91768 & 1.44  &	8.33 &183229	 &	1.40 & 8.04	\\
\hline
%
\multicolumn{14}{l}{%
  \begin{minipage}{15.5cm}%
{\vspace{1mm} \footnotesize {Superscript $^{\ast \ast}$ indicates distributions for which the power law is not ruled out at $p\le 0.1$ (conservative) level. Superscript $^{\ast}$ indicates distributions for which the power law is not ruled out at $p\le 0.05$ (lenient) level.  Abbreviations: patches -- number of area patches (number of population patches is slightly different), Salt Lake C. -- Salt lake City, S. Francisco -- San Francisco, Virginia B. -- Virginia Beach.} }
  \end{minipage}%
}
\end{tabular}}
\label{exponents}
%
\end{table*}

\begin{table*}[]
\caption{\textbf{{Exponents of power laws fitted to single-type distributions}}}
{\footnotesize \vspace*{3mm} \hspace*{2mm}
\begin{tabular}{p{1.7cm}|p{0.7cm} p{0.7cm} p{0.7cm} p{0.7cm} p{0.7cm} p{0.7cm} |  p{0.7cm} p{0.7cm} p{0.7cm} p{0.7cm} p{0.7cm} p{0.7cm}}
\hline
 & \multicolumn{6}{c|}{1990} &  \multicolumn{6}{c}{2010}  \\
type & patches & $\alpha_{a}$ & $R_{a}$ & patches & $\alpha_{p}$ & $R_{p}$ & patches & $\alpha_{a}$  & $R_{a}$  & patches\\
\hline
Whites MW & 12487 & 1.49(1) & 4.3 & 11059 & 1.54(1) & 4.9 & 21627  & 1.68(0) & 1.6 & 15730 & 1.74(1) & 2.7 \\
Whites BiW & 20319 & 1.73(1) & 2.0 & 17722 & 1.72(1) & 6.6 & 33403  & 1.67(0) & 2.2& 26924 & 1.69(0) & 8.0 \\
Blacks WB & 2811 & 1.55(1) & 3.4 & 2861 & 1.56(1) & 4.3 & 3761  & 1.64(1) & 1.9 & 3521 & 1.64(1) & 3.3 \\
Blacks BiB & 5454 & 1.77(1) & 3.2 & 5781 & 1.75(1) & 6.9 & 9158  & 1.76(1) & 2.0 & 9068 & 1.77(1) & 6.1 \\
Hispanics MH & 1886 & 1.82(2) & 1.1$^{\ast \ast}$ & 2213 & 1.71(1) & 3.4 & 4237  & 1.80(1) & 1.2$^{\ast \ast}$ & 5098 & 1.74(1) & 4.3 \\
Hispanics BiH & 5271 & 1.81(1) & 2.2 & 5933 & 1.71(1) & 6.0 & 12053  & 1.76(1) & 1.4 & 12732 & 1.73(1) & 5.8 \\
\hline
\multicolumn{13}{l}{%
  \begin{minipage}{15.5cm}%
{\vspace{1mm} \footnotesize {Abbreviations: LD -- low diversity, MD -- medium diversity  } }
  \end{minipage}%
}
\end{tabular}}
\label{exponents}
%
\end{table*}

Our test reveals that in $\sim$50\% of cities the hypothesis that patch areas are distributed according to the power law is statistically ruled out. This number is even larger for distributions of patch populations. It is important to remember that this pertains to a pure power law distribution, something we would not expect to find anyway due to data limitations and a large variety of different factors contributing to population dynamics. Note large values of $R$ for distributions constructed from the sample-wide pool of patches for which the visual fit of the power law to the data is excellent (see Fig.~2). This is because the $D \rightarrow 0$ when the number of samples drawn from the tested distribution increases  \citep{Tucker1959}. This means that the value of $\langle D \rangle$ is very small for cities with a large number of patches, and especially small for the pool of patches from all cities. This causes the values of $R$ to be large even if values of $D^{\star}$ are small. The power law model fits well to the data but not well enough to pass the test. In addition, note that, with some notable exceptions, the values of $R$ for distributions ruled out as pure power law are relatively small. Thus, such distributions are quite similar to power laws, and, for briefness, we are going to refer to all distributions in our sample as ``power laws" in the stand-in sense.

As can be seen in the bottom row of Fig.~2 the shapes of single-type patch size distributions depart from shapes of all-types patch size distributions. These departures are larger in 1990 than in 2010. Departures of single-type patch population distributions from the all-patches distribution are especially small in 2010. Table 3 summarizes values of power law exponents fitted to the empirical distribution of single-type patch area ($\alpha_a$) and patch population ($\alpha_p$) polled from all cities. Racial types listed in Table 3 include MW, BiW, MB, BiB, MH, and BiH. This table also summarizes the results of testing the power law assumption for these distributions. According to the test, only area distributions for MB type cannot be ruled out as the pure power law. Like in the case of all-patches distributions, single-type area distributions have smaller values of $R$ (are better fits to a power law model) than single-type population distributions.

\subsection{Results of multifractal analysis}
For each of the 41 urban areas, we calculated Renyi's generalized dimensions of racial patterns. We utilized the globally normalized \citep{Anderson2014} ``enlarged box-counting'' method \citep{Pastor-Satorras1996} with a dyadic sequence of the box sizes $\epsilon$. We calculate values of $D_q$ for $q$ between -20 and 20 in step of 1, but only keep the values fulfilling the following criteria, (1) the resultant value is in the range  ${0 \leq D_q \leq 2}$, (2) $R^2 \geq 0.9$ in the regression estimation of linear fit necessary to translate box counts into an estimate of $D_q$ \citep{Murcio2015}.

We have found that scaling condition (see above) was fulfilled for most of the urban areas for $q\ge 0$. We have also found that in all cases the scaling did not depend on the value of $q$ indicating that racial patterns are monofractal and can be characterized by the fractal dimension $D_0$ alone. This is not surprising inasmuch as racial patterns are reminiscent of land cover patterns (they consist of categorically-labeled cells in a raster), and monofractal patterns of urban land cover were reported in \citep{Chen2013,Kia2020}. 

Table 4 lists the values of fractal dimensions in 1990 racial patterns present in urban areas belonging to our dataset. Four different patterns were analyzed, a pattern formed by all patches ($D_0=D_{\rm ALL}$), and patterns formed by racial types MW, MB, and MH ($D_0=D_{\rm MW},D_{\rm MB},D_{\rm MH}$, respectively). The change of the value of the fractal dimension between 1990 and 2010 is indicated by symbols $\shortuparrow$, $\shortdownarrow$ and, - an increase, decrease, or no change. Urban areas/racial types for which we have insufficient data, or for which our analysis indicated nonfractal pattern, are labeled NA.

\begin{table*}[]
\centering
\caption{\textbf{{Fractal Dimensions}}}
{\footnotesize \vspace*{3mm} \hspace*{2mm}
\begin{tabular}{p{1.6cm}|p{0.73cm} p{0.73cm} p{0.73cm} p{0.73cm} || p{1.6cm}| p{0.73cm} p{0.73cm} p{0.73cm} p{0.73cm}}
\hline
city name & $D_{\rm ALL}$ & $D_{\rm MW}$ & $D_{\rm MB}$ & $D_{\rm MH}$ & city name & $D_{\rm ALL}$ &$D_{\rm MW}$ & $D_{\rm MB}$ & $D_{\rm MH}$ \\
\hline
Atlanta & 1.87 - & 1.83$\shortdownarrow$ & 1.50$\shortuparrow$ & NA  & New York  & 1.79$\shortdownarrow$ & 1.75$\shortdownarrow$ & 1.18$\shortdownarrow$ & 0.97$\shortdownarrow$\\
Austin & 1.76 - & 1.68$\shortdownarrow$ & 1.13$\shortdownarrow$ & 1.05$\shortuparrow$ & Orlando & 1.81$\shortdownarrow$ & 1.74$\shortdownarrow$ & 1.31$\shortdownarrow$ & 0.65$\shortuparrow$ \\
Baltimore & 1.83$\shortdownarrow$ & 1.77$\shortdownarrow$ & 1.43 - & NA  & Philadelphia & 1.86 - & 1.84$\shortdownarrow$ & 1.30$\shortdownarrow$ & 0.79$\shortdownarrow$\\
Boston & 1.82 - & 1.81$\shortdownarrow$ & 1.06$\shortdownarrow$ & 0.75$\shortdownarrow$ & Phoenix & 1.82$\shortuparrow$ & 1.78$\shortdownarrow$ & 1.05$\shortdownarrow$ & 1.30$\shortuparrow$\\
Charlotte &1.80- &	1.76$\shortdownarrow$ &1.31$\shortdownarrow$ &NA &  Pittsburgh &1.78 - &1.77$\shortdownarrow$ &	1.16$\shortdownarrow$& NA  \\
Chicago & 1.83$\shortdownarrow$ &	1.80$\shortdownarrow$ &	1.45$\shortdownarrow$ &	1.02$\shortuparrow$ & Portland & 1.81 - &	1.80$\shortdownarrow$ &	0.89$\shortdownarrow$ &	0.71$\shortuparrow$ \\
Cincinnati & 1.85$\shortdownarrow$ &	1.83$\shortdownarrow$ &	1.38$\shortdownarrow$ &	NA & Providence &	1.80- &	1.79$\shortdownarrow$ &	0.72$\shortdownarrow$ &	0.62$\shortuparrow$\\
Cleveland &	1.80 - &	1.78$\shortdownarrow$ &	1.41$\shortuparrow$ & 0.67$\shortuparrow$ & Riverside & 1.80 - &	1.55$\shortdownarrow$ &	0.99$\shortdownarrow$ &	1.16$\shortuparrow$ \\
Columbus &	1.85$\shortdownarrow$ &	1.83$\shortdownarrow$ &	1.44$\shortdownarrow$ &	NA & Sacramento &	1.83$\shortdownarrow$ &	1.75$\shortdownarrow$ &	0.82$\shortdownarrow$ &	0.9$\shortdownarrow$ \\
Dallas &	1.81 - &	1.74$\shortdownarrow$ &	1.36$\shortdownarrow$ &	1.14$\shortdownarrow$ & St. Louis &	1.81$\shortdownarrow$ &	1.79$\shortdownarrow$ &	1.37$\shortuparrow$ & NA\\
Denver &	1.80$\shortuparrow$ &	1.76$\shortdownarrow$ &	1.25$\shortdownarrow$ &	1.08$\shortuparrow$ & Salt Lake C. & 1.78 - &	1.77$\shortdownarrow$ &	NA &	0.97$\shortuparrow$  \\
Detroit & 1.86$\shortdownarrow$ &	1.83$\shortdownarrow$ &	1.58 - &	0.71$\shortuparrow$ & San Antonio &	1.83$\shortdownarrow$ &	1.65$\shortdownarrow$ &	1.21$\shortdownarrow$ &	1.53$\shortuparrow$ \\
Houston & 1.86$\shortdownarrow$ &	1.72$\shortdownarrow$ &	1.45$\shortdownarrow$ &	1.31$\shortuparrow$ & San Diego & 1.77$\shortdownarrow$ &	1.66$\shortdownarrow$ &	0.96$\shortdownarrow$ &	1.16$\shortuparrow$\\
Indianapolis &	1.81 - &	1.79$\shortdownarrow$ &	1.50$\shortdownarrow$ &	NA & S. Francisco & 1.71 - &	1.55$\shortdownarrow$ &	1.19$\shortdownarrow$ &	0.99$\shortdownarrow$ \\
Jacksonville &	1.83 - &	1.79$\shortdownarrow$ &	1.44 - &	0.42$\shortuparrow$ & San Jose &	1.82 - &	1.65$\shortdownarrow$ &	0.48$\shortuparrow$ &	1.17$\shortuparrow$\\
Kansas City &	1.82$\shortdownarrow$ &	1.79$\shortdownarrow$ &	1.40$\shortdownarrow$ &	0.95$\shortdownarrow$  & Seattle &	1.80$\shortdownarrow$ &	1.77$\shortdownarrow$ &	0.85$\shortdownarrow$ &	0.55$\shortuparrow$\\
Las Vegas &	1.83 - &	1.79$\shortdownarrow$ &	1.31$\shortdownarrow$ &	0.71$\shortuparrow$ & Tampa &	1.84 - &	1.80$\shortdownarrow$ &	1.26$\shortdownarrow$ &	1.06$\shortuparrow$\\
Los Angeles &	1.83$\shortdownarrow$ &	1.62$\shortdownarrow$ &	1.34$\shortdownarrow$ &	1.45$\shortuparrow$ & Virginia B. &	1.79$\shortdownarrow$ &	1.68$\shortdownarrow$ &	1.42$\shortdownarrow$ &	NA\\
Memphis & 1.82$\shortdownarrow$ &	1.75$\shortdownarrow$ &	1.55$\shortuparrow$ &	NA & Washington &	1.82$\shortdownarrow$ &	1.72$\shortdownarrow$ &	1.40$\shortuparrow$ &	0.84$\shortdownarrow$\\
Miami &	1.69$\shortuparrow$ &	1.61$\shortdownarrow$ &	1.21$\shortdownarrow$ &	1.27$\shortuparrow$ & Milwaukee & 1.82 - &	1.81$\shortdownarrow$ &	1.41$\shortdownarrow$ &	0.85$\shortdownarrow$\\
Minneapolis &	1.86 - &	1.86$\shortdownarrow$ &	0.99$\shortdownarrow$ &	0.5$\shortuparrow$ &  & &	 &	 &		\\
\hline
average 1990 &1.81	 &	1.75 &	1.24 & 0.94& average 2010 & 1.81	&1.63 & 1.25 &1.02	\\
 &	$\pm$0.037 &	$\pm$0.075	 &	$\pm$0.24 &	$\pm$0.28 &  &$\pm$0.035 &	$\pm$0.12 &		$\pm$0.30 & $\pm$0.029	\\
\hline
\multicolumn{10}{l}{%
  \begin{minipage}{12.5cm}%
{\vspace{1mm} \footnotesize {Values of fractal dimensions of spatial patterns formed by all cells of racial grid having a specified type; ALL--all types, MW--mono-racial Whites, MB--mono-racial Blacks, MH--mono-racial Hispanics. Fractal dimensions correspond to 1990 patterns; symbols $\shortuparrow$, $\shortdownarrow$ and, - indicate increase, decrease, or no change in the value of fractal dimension from 1990 to 2010. Abbreviations: Salt Lake C. -- Salt lake City, S. Francisco -- San Francisco, Virginia B. -- Virginia Beach.  } }
  \end{minipage}%
}
\end{tabular}}
\label{exponents}
%
\end{table*}

In our context, the fractal dimension is a measure of complexity that indicates the degree to which an urban form fills the 2D space. Examination of Table 4 leads to the following observations. (1) For any given city the value of $D_{\rm ALL}$ is always larger than values of fractal dimensions associated with individual racial type patterns. This is because racial type patterns are subsets of the entire urban pattern so they are expected to fill less underlying space. There has been no clear trend in the change of the value of $D_{\rm ALL}$ during the 1990-2010 period; the average value has not changed. (2) The fact that $D_{\rm MW}\lessapprox D_{\rm ALL}$ for most cities indicates that a 1990 spatial form associated with predominantly White inhabitants dominated the entire urban form. However, all values of $D_{\rm MW}$ decreased during the 1990-2010 period indicating a fragmentation of the MW form and its departure from the form of the entire urban area. (3) The fact that $D_{\rm MB} < D_{\rm ALL}$ for all cities indicates that a 1990 spatial form associated with predominantly Black inhabitants is less filling (more fragmented) than the entire urban form. There has been no clear trend in the change of the value of $D_{\rm MB}$ during the 1990-2010 period; the average value remained about the same.
(4)  The fact that $D_{\rm MH}\ll D_{\rm ALL}$ for most cities (Los Angeles and San Antonio are exceptions) indicates that a 1990 spatial form associated with predominantly Hispanic inhabitants consisted of several small enclaves. However, most values of $D_{\rm MW}$ increased during the 1990-2010 period indicating aggregation of spatial form associated with the predominantly Hispanic population.

\section{Discussion and Conclusions}
We analyzed the scaling and fractal properties of racial segregation patterns in American cities. The results of our analysis indicate that patterns of racial segregation are indeed fractal and that sizes of elements of racial patterns (patches of racial type) 
have heavy-tailed distributions which are either power laws or closely resemble the
power law. Overall, we demonstrated that racial patterns are previously unreported complex urban structures.

Before discussing the implications of this finding we first summarize the possible limitation of our analysis. (1) To perform our analysis we needed to use high-resolution racial grids. Such data represent streets and roads as uninhabited areas. Using image processing techniques we have removed streets, but major roads cannot be removed without producing artifacts. As a result, some large racial enclaves may have been divided affecting sizes of largest patches and thus tails of empirical distributions. However, values of exponents in fitted models are not sensitive to the values in tails of empirical distributions. (2) We use racial data restricted to fixed boundaries of census-defined urban areas. In \citep{Arcaute2015} it was showed that the values of scaling exponents in many urban systems may depend on the definition of urban boundaries; this effect was not investigated in this paper.
(3) We fit data to the assumed power law function and we quantify distributions in terms of the power law exponent even if the distribution is only an approximate power law. Such a procedure is widespread in studies in which scaling in networks (systems) is reported \citep{Broido2019}. However, whereas previous studies did not address a possibility that derived distribution may diverge from a true power law, we tested the power law hypothesis. Our test reveals, that in many cases, the power law hypothesis is statistically rejected (see Tables 2 and 3) and fitted models are only approximations of the empirical distributions. We argue that these approximations are sufficiently good to not change our conclusions and to not diminish the significance of our findings.

\begin{figure*}[t]
	\includegraphics[width=15cm]{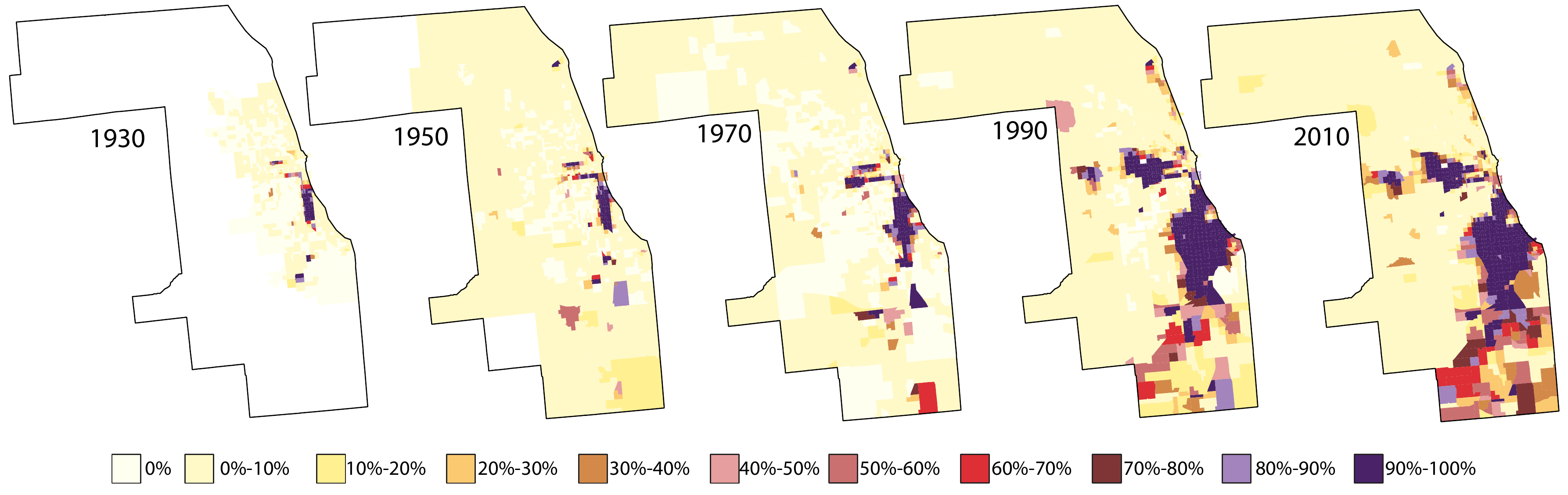}
\caption{Evolution of distribution of Black subpopulation in Chicago from 1930 to 2010. Different colors indicate different shares of Blacks among local inhabitants.
}
\label{fig3}
\end{figure*}

The significance of our findings is that they provide observational constraints on models of racial segregation. To discuss these constraints we need to give a short background on how racial segregation patterns arose in American cities. At the beginning of the twentieth-century major American cities were monoracial White (type MW). During the 1916-1970 period, the migration of Blacks from rural South to major cities make them more racially diverse as a whole. However, these Black migrants did not integrate with Whites, instead, they formed their own patches (type MB). Fig.~3 illustrates this point using Chicago, IL as an example. This figure clearly shows that segregation was present in Chicago from the moment the city ceased to be monoracial. Initial MB patches persisted and grew, and new MB patches occasionally formed and began their own growth, eventually merging with older patches. A similar process occurred more recently (from the 1960s to present) with immigrants from Latin America and Asia. The present racial pattern is the result of complex dynamics illustrated in Fig.~3, this dynamics is typical for all major American cities \citep{DmowskaSS2020}.

Given our results, an evolutionary model of the process responsible for observed patterns of segregation in American cities must lead to a fractal pattern with scaling properties of patch size. The most popular model of racial segregation is the Schelling model \citep{Schelling1971}. Schelling's model is an agent-based model that demonstrates how a macrobehavior (racial segregation) emerges from micromotives (individual preferences of inhabitants). It has been very popular because the transformation from integration (initial condition) to segregation (final configuration) occurs even when individual preferences are only mildly discriminatory; its an example of an invisible-hand explanation of the result of a complex process. It is important to note that racial patterns produced by Schelling's model are neither fractal nor scalable (see, for example, \citep{Clark2008}). Thus, the Schilling model cannot explain the racial patterns observed in the US. On the other hand, \citep{Hatna2012} reported that the modified Schelling model produces patterns that can be related to observed racial patterns in two cities in Israel.

The explanation of this discrepancy is in fundamentally different histories of societies in the US and in Israel. In Israel, three major population groups (Jews, Arabs, and Christians) lived in an integrated society for centuries before they started to segregate in the twentieth century due to changing attitudes -- this is the scenario that the Schelling model is designed for (evolution from integration to segregation). On the other hand, in the US, the population in major cities was never integrated, so an evolutionary model aiming at explaining segregation observed today must start from a monoracial (White) city to which people of different races are gradually added (see Fig.~3). This calls for a dynamical process very different from what Schelling's model encapsulates. 

We propose a hypothesis on dynamics leading to segregation patterns observed in US cities. This hypothesis accounts for what is presently observed (fractal form and scaling of racial patches) and also for the series of racial maps showing the evolution of segregation pattern (Fig.~3). Without the loss of generality, we will describe our hypothesis using two racial population types, W (majority) and B (minority). The initial condition is the urban area inhabited only by W. Population B migrates to the city and establishes its residence on top of the existing urban structure. We propose that the dynamical principle behind the pattern formed by residences of population B is the growth by preferential attachment (see \citep{Perc2014} for a review of preferential attachment). In our context, the growth involves three fundamental processes: nucleation, growth, and linking \citep{Small2015}. When a new B resident is added to the city, his residence can be isolated from existing B patches (nucleation), attached to an already existing B patch (growth), or connect two already existing B patches (linking). The principle of preferential attachment states that the new B resident will most likely attach to preexisting B patches with a probability linearly increasing with the population count of such patch. This is a very plausible assumption because new migrants tend to settle in neighborhoods when they will get the most support. Over time, such dynamics will create B patches with a power law distribution of sizes \citep{Fu2005,Yamasaki2006} which is what we have found in US cities. The fractal form of the B pattern is not related directly to racial segregation, instead, it is related to the fact that cities as a whole grow to form a fractal pattern \citep{Makse1995,Murcio2011} of which the B pattern is a subset. The complement of the B pattern is the A pattern which is also fractal for the same reason. 

A simple, analytical model of growth by preferential attachment \citep{Fu2005,Yamasaki2006} displays the following features. (1) in the long time limit, the size distribution of patches is a power law
\begin{equation}
p(x) \propto x^{-1 -\frac{1}{1-b}}
\end{equation}
\noindent (2) The value of parameter $b$ dictates the strength of preferential attachment; smaller values of $b$ result in larger preferential attachment, the minimum value of the exponent is 2 (for $b=0$). (3) The size of a patch is proportional to its age. (4) the power law steepens with decreasing magnitude of preferential attachment (increased value of parameter $b$).

Are those features compatible with our findings? (1) We found that indeed size distributions are power laws. (2) We found that exponents of those power laws to be
smaller than 2, whereas the simple model predicts values of exponent $\ge 2$. (3) The evolutionary series of racial maps in Chicago (Fig.~3) suggests that indeed the largest patch in 2010 is a descendant of the oldest patch (the largest patch in 1930).
The smallest patches in 2010 are the youngest. (4) We found that in 33 out of 41 cities the value of exponent increased from 1990 to 2010 resulting in patch distribution with relatively more small patches which can be interpreted as a decrease in segregation. In terms of the model, larger exponent means smaller preferential attachment, which is consistent with the change in micromotives leading to larger integration.

Overall, the only disagreement between the simple model based on preferential attachment and our findings is the value of the exponent. However, the simple model assumes a constant strength of preferential attachment and does not account for intra-city migration. Taking these factors into account may lower the value of the exponent. Future research is needed to translate our hypothesis into a testable model.

\vspace*{2mm}
{\noindent {\bf Acknowledgments.} This work was supported by the University of Cincinnati Space Exploration Institute.}


\end{document}